\documentclass[twocolumn, prl, preprintnumbers, showpacs, amsmath, amssymb, superscriptaddress]{revtex4-1}
\usepackage{graphicx}
\usepackage{amsmath}
\usepackage{physics}
\usepackage{bm}
\usepackage[english]{babel}
\usepackage{xr}
\newcommand{\ee}{\mathrm{e}}
\renewcommand{\rm}{\mathrm}
\allowdisplaybreaks

\begin{document}
\title{Magnetism Induced by Periodically Driven Non-Magnetic Impurities on Surfaces with Spin-Orbit Coupling}

\author{M. Etxeberria-Etxaniz}
\affiliation{Departamento de Física, Universidad del País Vasco UPV/EHU, 48080 Leioa, Spain}

\author{A. Arnau}
\affiliation{Departamento de Polímeros y Materiales Avanzados: Física, Química y Tecnología, Universidad
del País Vasco UPV/EHU, 20018 Donostia-San Sebastián, Spain}
\affiliation{Centro de Física de Materiales (CFM/MPC) CSIC-UPV/EHU, 20018 Donostia-San Sebastián, Spain}
\affiliation{Donostia International Physics Center (DIPC), 20018 Donostia-San Sebastián, Spain}

\author{A. Eiguren}
\affiliation{Departamento de Física, Universidad del País Vasco UPV/EHU, 48080 Leioa, Spain}
\affiliation{Donostia International Physics Center (DIPC), 20018 Donostia-San Sebastián, Spain}
\affiliation{EHU Quantum Center, Universidad del País Vasco UPV/EHU, 48080 Leioa, Spain}

\begin{abstract}
We investigate the response of the Rashba spin-orbit system to a time-periodic scalar potential, in order to determine whether an induced magnetization exists. We approach this by employing the Floquet-Green function method within the Keldysh formalism, computing the non-equilibrium steady state of the system.
We find that, even in the absence of an external magnetic field, the system evolves into a state with an oscillating magnetization density that is remarkably rich in structure.
We provide a detailed physical interpretation of the results by performing a Fourier decomposition in non-local momentum-space, which helps to uncover the physical origin of the induced magnetic field in terms of Fermi surface spin polarization and the system's dynamical character.
\end{abstract}

\maketitle

Spin-orbit coupling (SOC) is the lowest-order term in the relativistic regime that lifts spin degeneracy while preserving time-reversal symmetry. Therefore, it is present even in non-magnetic materials, particularly those which contain heavy elements, where it appears as an intrinsic effect. Since the photoemission experiment by LaShell and co-workers \cite{LaShell1996} revealed energy splitting in the Au(111) surface state due to SOC, it has been recognized as a source of a wide variety of phenomena in condensed-matter physics. Notably, a new field has emerged based on the realization that SOC can lead to topological insulating electronic phases \cite{HasanKane2010}. Surfaces are even more important in this context because it is precisely the lack of inversion symmetry that is responsible for lifting Kramers degeneracy \cite{rashba-bychkov1984_1}. Moreover, surfaces are directly accessible to experimental techniques such as angle-resolved photoemission spectroscopy (ARPES), scanning tunneling microscopy (STM), and other surface-sensitive methods~\cite{Damascelli2003, Binnig1982}. In this context, SOC offers the possibility of manipulating the spin without the use of any external magnetic fields and, hence, the Rashba effect is an excellent candidate for spintronic applications \cite{datta-das1990, marchenko2012, noel2020}. Even in light elements, like carbon, the Rashba effect could give rise to spin polarized currents in chiral nanotubes \cite{PhysRevLett.93.176402}. Additionally, the Rashba coupling can be enhanced by the application of an external electric field~\cite{PhysRevB.93.165424}. 

On surfaces, impurities lead to an even more pronounced symmetry reduction by breaking the remaining translational invariance. In the non-relativistic limit, and for a free-electron-like system with a spherical Fermi surface, static impurity perturbations give rise to charge density oscillations with a wavelength corresponding to a momentum transfer equal to twice the Fermi momentum. These are known as Friedel oscillations \cite{Friedel1958}. The presence of spin polarization on the Fermi surface makes the problem richer and more complex, as the scattering matrices become spin-dependent. For instance, back-scattering is forbidden in the presence of spin-orbit coupling, whereas in the non-relativistic case, it is the dominant mechanism. Nevertheless, STM experiments measuring quasiparticle interference patterns reveal this physics through the role of `static' impurities~\cite{Eiguren2011} on a metal surface. It has also been shown that magnetic impurities can induce Friedel oscillations with skyrmion-like spin textures~\cite{lounis2012, bouaziz2018}.

In this letter, we investigate the dynamics of impurities on the simplest relativistic surfaces, described by the Rashba model, aiming at determining whether a non-magnetic, localized and time-periodic perturbation can induce a magnetic response in a non-magnetic surface with intrinsic SOC. It was recently demonstrated that phonons may induce an oscillating magnetization density even in non-magnetic materials, by breaking spatial translational invariance and unbalancing the spin population across the Brillouin zone~\cite{ggurtubay2020}. Similarly, our idea here is that breaking time-translational invariance should introduce magnetic oscillations in both time and space domains. As a proof of concept for this phenomenon, we choose a periodically oscillating, cylindrical-shaped perturbation to mimic a vibrating adatom or molecule on a surface. Despite the toy-model nature of the system, the resulting behavior is rich and complex, as described shortly.

We adopt the Floquet-Green function method (FGFM) \cite{tsuji2008, aoki2014}, which provides a theoretical framework for studying periodically driven systems out of equilibrium using Green’s function formalism. As it is common in most non-equilibrium systems, the time dependence is inherently non-local. However, instead of using the time variables $t$ and $t'$, it is physically more natural to express the dynamics in terms of their relative $t_\rm{rel} \equiv t - t'$ and average $t_\rm{avg} \equiv (t+t')/2$ values. Thus, we write the Green's function as 

\begin{equation} \label{wigner_repr}
\begin{split}
G_n (\omega) = \frac{1}{T} &\int_{-T/2}^{T/2} \dd t_\rm{avg} \: \ee^{in \omega_0 t_\rm{avg}} \\
\times& \int_{-\infty}^{\infty} \dd t_\rm{rel} \: \ee^{i\omega t_\rm{rel}} \,G(t_\rm{rel}, t_\rm{avg}),
\end{split}
\end{equation}
where $\omega_ 0 \equiv 2\pi / T$ is the fundamental frequency. 
The resulting function is referred to as the \emph{Wigner representation}, omitting the spatial coordinates and spin indices for brevity. 
The matrix form of the Green's function is defined as
\begin{equation} \label{floquet_repr}
G_{mn} (\omega) \equiv G_{m-n} \left( \omega + \frac{m+n}{2} \omega_0 \right).
\end{equation}
This is the so-called \emph{Floquet representation} of the Green’s function. One of the key aspects of the above expression is that the association of $\omega$ with $\omega + (m+n)\omega_0 /2$ requires restricting the former to $\omega \in [ -\omega_0 /2,\; \omega_0 /2)$
\cite{sambe1973, gavensky2018}. This introduces a frequency Brillouin zone, defining frequency equivalence classes related by the addition of $ \omega_0$, analogous to the momentum-space Brillouin zone resulting from Bloch's theorem.

Formally, the Dyson equation is given by
\begin{equation} \label{dyson}
\begin{aligned}
		G_{mn}&(\bm{r}, \bm{r}', \omega) = G^0_{mn}(\bm{r}, \bm{r}', \omega) \\
		&+ \int \dd\bm{r}'' \sum_{m'n'} G^0_{m m'}(\bm{r}, \bm{r}'', \omega) V_{m'n'}(\bm{r}'') G_{n'n}(\bm{r}'', \bm{r}', \omega),
\end{aligned}
\end{equation}
where, for the moment, we include the non-local spatial variables, but we will omit them where relevant for brevity. 
$G^0$ denotes the unperturbed Green's function and $V$ the external perturbation. Spin indices are again omitted, but they enter in the same way as the $m$ and $n$ indices. In this way, non-local time integrations are transformed into convolution formulae as multiplications in frequency space, which is numerically convenient, and Green’s functions reduce to matrices of finite dimension \cite{tsuji2008}. Equation \ref{dyson} is aimed at describing the external drive on the electrons by a time-dependent, local potential \( v(\bm{r},t)\, \delta(\bm{r} - \bm{r}') \), but it needs to be extended to the Keldysh formalism in order to account for the non-equilibrium nature of the problem. Formally, the retarded ($G^\rm{R}$), advanced ($G^\rm{A}$), and the so-called Keldysh ($G^\rm{K}$) Green's functions are needed to describe the system. The so-called lesser Green's function $G^<$, defined as $G^< = \frac{1}{2}(G^\rm{K} - G^\rm{R} + G^\rm{A})$ (see Sec. I of Supplemental Material \cite{suppmat}), is essentially playing the role of the spectral density, analogous to the imaginary part of the Green's function in equilibrium, that leads to the density of occupied states~\cite{aoki2014}.

Now, our objective is to study the time dependence of the charge and magnetic response of the system, which is accessible by going back to the time domain as

\begin{equation} \label{antitransf}
\resizebox{1\hsize}{!}{$
\displaystyle G^<(t,t') = \sum_{mn} \int_{-\omega_0 /2}^{\omega_0 /2}\, \frac{\dd \omega}{2\pi} \, \ee^{-i (\omega + m \omega_0)t } \ee^{i (\omega + n \omega_0) t'} G^{<}_{mn}(\omega).
$}
\end{equation}

In our case, the unperturbed system corresponds to the Rashba electron gas, and the Green’s function is given by
\begin{equation}
\resizebox{1\hsize}{!}{$
\bm{G^0} (\bm{r}, \bm{r}',\omega) = 
\begin{pmatrix}
G_\rm{d}^0 (\bm{r}, \bm{r}',\omega) & -\ee^{-i\theta_{\bm{r}\bm{r}'}} G_\rm{od}^0 (\bm{r}, \bm{r}',\omega) \\[5pt]
\ee^{i\theta_{\bm{r}\bm{r}'}} G_\rm{od}^0 (\bm{r}, \bm{r}',\omega) & G_\rm{d}^0 (\bm{r}, \bm{r}',\omega)
\end{pmatrix},
$}
\end{equation}
where we have introduced the Pauli formalism for the electron spin, in which Green’s functions become $2 \times 2$ spin-density matrices, denoted by bold symbols $\bm{G}$.
If the quantization axis is perpendicular to the surface plane, the off-diagonal elements represent the in-plane components of the spin polarization, while the diagonal elements represent the charge density, as no net magnetism is present in the system. In our notation, $\theta_{\bm{r}\bm{r}'}$ indicates the angle between $\bm{r}$ and $\bm{r}'$ points in space, and $G_\rm{d}^0$ and $G_\rm{od}^0$ are given by combinations of Hankel $H_{0}^{(1)}$~and $H_{1}^{(1)}$ functions, respectively (see Secs. II and III of Supplemental Material \cite{suppmat}). Due to the equilibrium nature of the unperturbed Green’s function, it does not depend on the variable $t_\rm{avg}$. Therefore, the Wigner representation has a single non-zero frequency component, $\bm{G}_{\bm{n}=\bm{0}}^{\bm{0}} (\bm{r}, \bm{r}',\omega) = \bm{G^0} (\bm{r}, \bm{r}',\omega)$. The next step requires writing the full Floquet representation of $\bm{G^0}$ using Equation \ref{floquet_repr}.

The following ingredient for the Dyson equation is the perturbation, which we choose to be a simple cylindrical potential centered at the origin and localized within a small radius $r_V$.
As already mentioned, our objective is primarily to study the possible magnetic response to a non-magnetic perturbation and, therefore, the impurity is a diagonal matrix in the $2 \times 2$ spin representation. We also assume that the time dependence is sinusoidal with frequency $\omega_0$. Altogether, the spin-diagonal component of the potential is modeled as
$V(\bm{r},t)=v(\bm{r})\cos(\omega_0 t)$,
where $v(\bm{r})$ is a localized potential; we consider a cylindrical shape $v(\bm{r})=v_0\,\Theta(r_V-r)$, with $\Theta$ the Heaviside function, without loss of generality. Its Floquet representation is
$ V_{mn}(\bm{r})=\frac{1}{T}\int_{-T/2}^{T/2}\!\dd t\, \ee^{i m\omega_0 t} V(\bm{r},t)\,\ee^{-i n\omega_0 t}. $

The physical intuition driving our investigation is that an oscillating scalar potential as described above is, up to a gauge transformation, equivalently represented by the pure-gauge vector potential $\mathbf A(\bm r,t)=\nabla v(\bm r)\sin(\omega_0 t)/\omega_0$, so we have $\mathbf E=-\partial_t\mathbf A=-\nabla v(\bm r)\cos(\omega_0 t)$ and zero magnetic field $\mathbf B=\nabla\times\mathbf A=0$ by construction. To see how this impurity field couples to the Rashba mode, considering minimal coupling $\bm p\!\to\!\bm p-e\mathbf A$, it is easy to see that the Rashba term $H_\rm{R}=\alpha_\rm{R}(\boldsymbol{\sigma}\times\bm p)_z$ acquires, to linear order in $\mathbf A$, the azimuthal, purely spin-driving coupling,
\begin{align}
	\delta H_\rm{R}(t) = 
	e(r,t)\,\boldsymbol{\sigma}\!\cdot\!\hat{\boldsymbol\phi} 
	= i e(r,t)\big(e^{i\theta}s_- - e^{-i\theta}s_+\big),
	\label{eq:azimuth}
\end{align}
where $\hat{\boldsymbol\phi}= - \hat{\bm{\imath}} \sin(\theta) + \hat{\bm{\jmath}} \cos(\theta)$ is the azimuthal unit vector,
$e(r,t)=\alpha v_0/\omega_0\, \cos(\omega_0 t)\,\delta(r{-}r_V)$
is the modulus of the time-oscillating electric field localized in a ring-shaped region at radius $r_V$, and
$s_\pm=\tfrac12(\sigma_x\!\pm i\sigma_y)$. 
Crucially, because the drive is purely azimuthal, it acts as an effective in-plane tangential field localized on the ring \(r=r_V\). Its localized character, in turn, breaks translational symmetry and generates long-range, Friedel-like spin oscillations with polarization along $\hat{\boldsymbol\phi}$.

At this point, we are ready to study the coupling of the Rashba electron gas and our model time-dependent impurity by means of the Dyson equation in Eq. \ref{dyson}. This is a sparse linear system and is solved directly using the \texttt{PARDISO} package \cite{pardiso}, considering a combined computational index comprising real-space, spin, and frequency variables to construct the matrices that represent the system. With this technique, we eventually obtain the lesser Green's function, from which we can compute the charge and spin density variations in space and time.
In particular, we are interested in the charge and magnetization densities which are accessible from the $\bm{G^{<}}$ matrix~as
\begin{equation}
\begin{split}
n (\bm{r}, t) &= -\frac{i}{2} \lim_{(\bm{r}',t') \rightarrow (\bm{r},t)} \Tr \sigma_{0} \bm{G^{<}} (\bm{r}, \bm{r}', t, t'), \label{dens_def} \\ 
m_{x,y,z} (\bm{r}, t) &= -\frac{i}{2} \lim_{(\bm{r}',t') \rightarrow (\bm{r},t)} \Tr \sigma_{x,y,z} \bm{G^{<}} (\bm{r}, \bm{r}', t, t'), 
\end{split}
\end{equation}
where the traces are calculated in spin space and $\sigma_{0,x,y,z}$ are the Pauli matrices.

\begin{figure}[t]
\includegraphics[width=1\columnwidth]{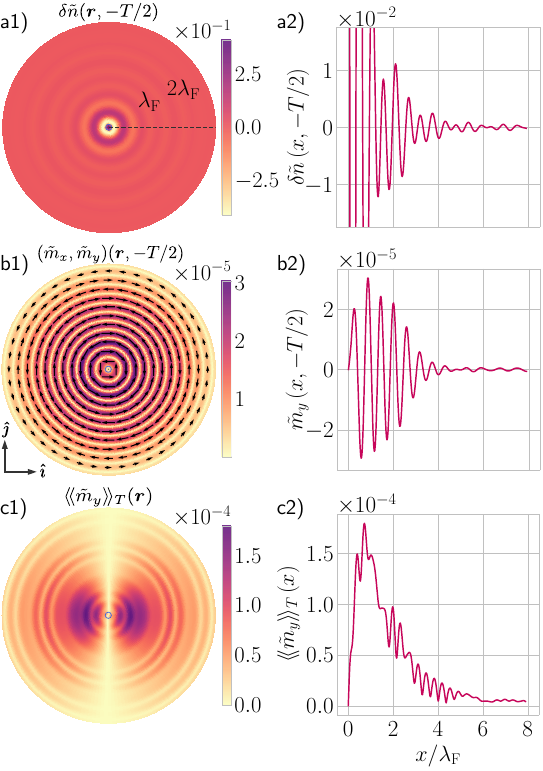}
\caption{ Induced charge density $\delta \tilde{n} \equiv (n-n_0)/n_0$ (a), in-plane magnetization components $(\tilde{m}_x, \tilde{m}_y) \equiv (m_x /n_0, m_y/ n_0)$ (b), and the standard deviation over one period
$\langle\!\langle \tilde{m}_y \rangle \!\rangle_T(\bm{r})$ (c).
All densities are normalized by the unperturbed value $n_0$ and evaluated at time $-T/2$. Spatial coordinates are expressed in units of the Fermi wavelength $\lambda_{\mathrm{F}}/2 = 2\pi/(k_{\mathrm{F}+} + k_{\mathrm{F}-})$. Panels (a1) and (a2) show, respectively, the two-dimensional distribution of the induced charge and its cross section along the positive $x$-axis (dashed line in the former). Panels (b1) and (b2) show the in-plane magnetization: (b1) displays its modulus and direction as arrows representing the normalized vector
$\tilde{m}_x \bm{\hat{\imath}} + \tilde{m}_y \bm{\hat{\jmath}}$, and (b2) shows its cross section along the positive $x$-axis. Panels (c1) and (c2) depict the time-averaged standard deviation $\langle\!\langle \tilde{m}_y \rangle \!\rangle_T(\bm{r})$ in the two-dimensional plane and its cross section, respectively. A small blue circle at the origin marks the location of the impurity. Panels (a1), (b1) and (c1) share a radial extension of $2.5 \lambda_\rm{F}$ in order to facilitate visualization. Likewise, the vertical scale in panel (a2) is reduced so that induced charge far from the impurity is noticeable. }
\label{dos_3x2}
\end{figure}

In Figure~\ref{dos_3x2}, we present the normalized induced charge and in-plane magnetization densities evaluated at the representative time $-T/2$. Panels (a1) and (b1) show the two-dimensional real-space distributions of the induced charge and magnetization, respectively. Panels (a2) and (b2) display the corresponding cross sections along the positive $x$-axis. First of all, we observe that, despite the non-magnetic nature of both the system and the external perturbation, the system responds with a time-dependent magnetization. We have verified that a static perturbation (\( \omega_0 \rightarrow 0 \)) gives rise to the usual Friedel oscillations, without inducing any magnetization as expected (not shown). Panels (b1) and (b2) show the magnetic profile at a given arbitrary time. However, the system evolves in such a way that it remains overall non-magnetic, in the sense that the averages over time and space are exactly zero. This is because the system evolves sinusoidally in time and because, in real-space, the points $\bm{r}$ and $-\bm{r}$ are connected to opposite magnetic moments. In panels (c1) and (c2) we plot the time average over one period of the mean square deviation of the $y$ component of the magnetic response, calculated as $\langle\!\langle \tilde{m}_y \rangle \!\rangle_T(\bm{r}) = \sqrt{ \langle \tilde{m}_y^2(\bm{r}, t) \rangle_T }$ since $\langle \tilde{m}_y(\bm{r}, t) \rangle_T$ is obviously zero, and where angle brackets denote time averages over one period $T$
(see Sec. IV of Supplemental Material~\cite{suppmat}).

Overall, for the chosen parameters, we find that the strength of the magnetic patterns is about three orders of magnitude weaker than the scalar charge patterns in this paradigmatic surface (see Sec. V of Supplemental Material \cite{suppmat} for a brief analysis of the induced magnetization dependence on the key drive parameters). However, many bismuth compounds and transition-metal dichalcogenide layers exhibit at least an order of magnitude stronger spin-orbit coupling, and one could therefore reasonably expect the induced magnetization to be only about two orders of magnitude smaller than the charge oscillations.

We observe, particularly in panel (b2) of Figure \ref{dos_3x2}, that the profile of the magnetic response exhibits a shorter-wavelength modulation enveloped by a longer wavelength component. This is not surprising, since there are two characteristic scales in the Fermi contours, associated with the Fermi wavevector $\bm{k_\rm{F}}$ and the Rashba parameter $\alpha_\rm{R}$. 

Motivated by the need to investigate oscillation scales and identify the possible scattering processes responsible for the emergence of magnetization, we performed a double Fourier decomposition in non-local momentum-space, which allows us to compute the lesser Green's function for initial and final momenta $\bm{k}$ and $\bm{k}'$. Based on Equations \ref{dens_def}, we define the following non-local momenta Green's functions for the magnetic components,
\begin{equation}
\begin{split}
G^{<}_{0,x,y,z}(\bm{k}, \bm{k}',t) &= \frac{1}{2} \lim_{t' \rightarrow t} \Tr \sigma_{0,x,y,z} \bm{G^<}(\bm{k}, \bm{k}',t,t').
\end{split}
\end{equation}
These quantities allow us to analyze the various scattering processes that give rise to both classical and magnetic Friedel oscillations. For example, by fixing $\bm{k}$ at a point close to the Fermi contours (i.e., $|\bm{k_{\rm{F}\pm}}|$) and plotting the resulting functions as a function of $\bm{k}'$, we can clearly identify the final states that contribute most significantly to the induced charge and magnetization densities.

\begin{figure}
\includegraphics[width=1\columnwidth]{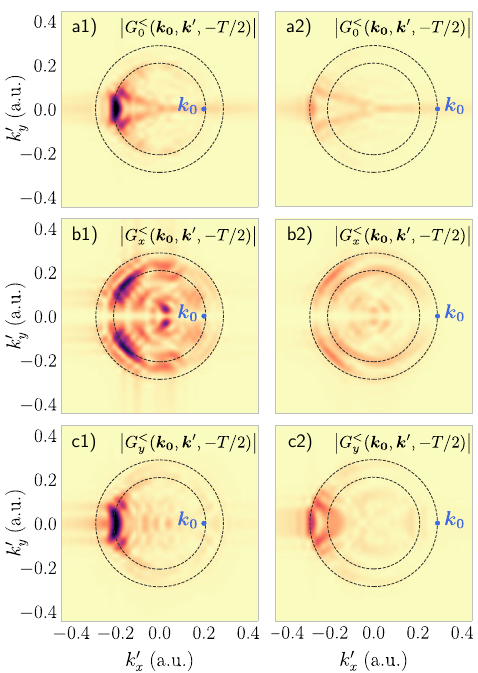}
\caption{
Absolute value of momentum resolved lesser Green's functions $G^{<}_0$ (a), $G^{<}_x$ (b) and $G^{<}_y$ (c) for initial momentum $\bm{k_0}$ close to the inner Fermi contour (left-hand side panels (a1), (b1) and (c1)) and the outer Fermi contour (right-hand side panels (a2), (b2) and (c2)), respectively. All the functions are evaluated at time $-T/2$, but the results at other snapshots are very similar (see Sec. VI of Supplemental Material \cite{suppmat}).}
\label{2d_ft}
\end{figure}

Panels (a1) and (a2) of Figure \ref{2d_ft} show that the induced charge density is dominated by the back-scattering process, with clear hotspots lying in the intraband back-scattering regions for initial momenta located on both the inner and outer Fermi contours. Quite obviously, this deviates from the standard picture because it would be impossible if the impurity were static, as the spins are opposite and the overlap would be negligible. In other words, this figure shows that the dynamical nature of the problem enables scattering processes that are forbidden in the static case. In contrast, panels (b) and (c), which are associated to the $x$ and $y$ components of the induced magnetization, demonstrate the non-trivial nature of the scattering patterns in momentum-space. In particular, panels (b1) and (b2)
show that both intraband and interband contributions are relevant for the induced magnetization, with significant scattering occurring also at intermediate angles. Likewise, it is interesting to note that the $x$ component of the induced magnetization displayed in the previous figures does not show any contribution from back-scattering channels but, instead, displays hotspots at an angle of approximately $\sim\!\!\pi/6$ around the back-scattering direction. There is no $x$ component of the spin in perfect back-scattering because of the circulating nature of the spin polarization. However, the density of final states increases as the scattering angle approaches $\pi$ in two dimensions. As a result, the hotspot is located at an intermediate angle, satisfying a compromise between these two requirements. These results also provide a consistent numerical check: the dominant contributions lie close to the Fermi contours, which is not imposed explicitly but rather handled by the Keldysh formalism. It is worth mentioning that our results are related to the Edelstein effect \cite{EDELSTEIN1990233} and spin pumping phenomena, which describe spin polarization and spin current generation under electric fields or dynamic magnetization. However, in contrast to these traditional setups, here we demonstrate that a purely time-periodic, scalar potential in a Rashba system can induce a magnetic response even in the absence of magnetic materials or spin injection. 

Regarding the possible experimental detection of this phenomenon, similar to ESR-STM techniques that probe spin dynamics with microwave excitation \cite{Choi2017}, the dynamical magnetization could be measured using spin-polarized STM \cite{Wiesendanger2009,Friedlein2019} operated under continuous wave microwave frequency modulation in the THz regime. The microwave frequency range should include the vibrational frequency of the impurity adatom or molecule. The electric field at the tunnel junction is enhanced by the presence of the STM tip and provokes the vibrational excitation \cite{Ho2023}. In such a configuration, the spin-polarized tunneling current depends on the local magnetization, yielding higher currents for parallel alignment between the tip and the sample magnetization; thus, even small variations in the induced magnetization could be resolved with magnetized tips oriented parallel to the surface, using a lock-in technique that detects the current changes.
\vspace{0.01cm}

In conclusion, we have shown that a periodically driven, scalar potential represents a new mechanism for dynamically generating spin polarization in systems with sizable spin-orbit coupling. The dynamic nature of the perturbing potential is essential and it affects both the induced charge and magnetization densities, that are determined by scattering processes absent in the static case. As a proof of concept and comprehensible system to model a vibrating adatom or molecule on a surface, we have considered the simplest Rashba surface perturbed by a localized, cylindrical-shaped and oscillating scalar potential, that has permitted us to rationalize our results. We find that both inter and intraband scattering processes involving spin-flip transitions are determinant to produce the induced magnetization pattern. This mechanism provides an alternative route to spin control in non-magnetic systems, extending the scope of non-equilibrium spintronics.

We thank Nacho Pascual and Sebastian Bergeret for useful discussions. This work was supported by grant IT-1527-22 from the Department of Education, Universities and Research of the Basque Government; predoctoral fellowship PIF21/163 from the University of the Basque Country;
Grants PID2022-137685NB-I00 (M.E.-E., A.A., A.E.) and PID2022-138210NB-I00 (A.A.), funded by MCIN/AEI/10.13039/501100011033 and by ”ERDF A way of making Europe”.

\textit{Data avalaibility}—The data that support the findings of this letter are openly available \cite{data}.

\bibliographystyle{apsrev}
\bibliography{biblio}

\onecolumngrid
\vspace{1cm}
\begin{center}
\textbf{End Matter}
\end{center}

\twocolumngrid

\textit{Appendix: Fourier transform of the 1D cross sections of the charge and magnetic response---} For completeness, in Figure 3 we include the one-dimensional cross sections of the charge and magnetic responses presented in panels (a2) and (b2) of Figure \ref{dos_3x2}. This analysis is motivated by the need to confirm the presence of two characteristic oscillation scales, associated with the Fermi wavevector $\bm{k_\mathrm{F}}$ and the Rashba parameter $\alpha_\mathrm{R}$. 

The results can be rationalized by taking into account that the Rashba coupling ($\alpha_{\mathrm{R}}$), the Fermi energy ($\varepsilon_{\mathrm{F}}$), and the effective mass ($m^*$) completely determine the model electron gas system. The outer and inner Fermi wavevectors are given by,
\begin{equation}
\frac{k_{\rm{F}\pm}}{m^*} = \pm \alpha_\rm{R} + \sqrt{\left( \alpha_{\rm{R}} \right)^{2} + 2\varepsilon_\rm{F}/m^*},
\end{equation}
where $m^*$ is the effective mass that we set as $m^*= 1\; \rm{a.u.}$ (see Sec. II of Supplemental Material \cite{suppmat}).

In panel (a) of Figure 3 we observe that the charge density is dominated by 
wavevector components of approximately $k_{\rm{F}+} + \, k_{\rm{F}-}$. 
Note that in the limit
$\alpha_\rm{R} m^* \ll k_\rm{F}$ 
we would have $k_{\rm{F}+} + \, k_{\rm{F}-} \approx 2k_\rm{F}$.
Although this approximation does not hold exactly in our system, it highlights that the maximum approximately at \( k_{\rm{F}+} + k_{\rm{F}-} \) is a signature of a back-scattering process at the Fermi level. Therefore, the resulting charge density oscillations can be interpreted as classical Friedel oscillations with half the Fermi wavelength $\lambda_\rm{F}$.
In contrast, panel (b) shows that the Fourier decomposition of the magnetization reveals two main wavevector components at somewhat smaller wavevector values. This suggests that the Fourier response displays a distributed $q$-spectrum rather than two back-scattering peaks at $\theta=\pi$; hence the simple $\pi$–scattering combinations of $k_{\mathrm F+}$ and $k_{\mathrm F-}$ are insufficient to account for the observed magnetization pattern. This motivated us to address the problem from a broader perspective, using 
non-local momentum Green’s functions 
\( G^{<}_{0,x,y,z}(\bm{k}, \bm{k}', t) \), 
as represented in Figure~\ref{2d_ft}.

\vspace{0.2cm}
\begin{figure}\label{1d_ft}
    \includegraphics[width=0.9\columnwidth]{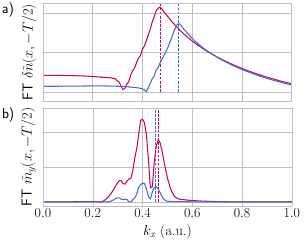}
    \caption{Figure 3. Fourier transform of the cross section of the induced charge (a) and the $y$ component of the induced magnetization (b) evaluated at time $-T/2$ (arbitrary units). 
Panel (a) shows the results for two representative values of the Fermi energy: $\varepsilon^{(1)}_\rm{F} = 0.03~\mathrm{a.u.}$ (magenta) and $\varepsilon^{(2)}_\rm{F} = 0.04~\mathrm{a.u.}$ (blue). Panel (b) presents data for two representative values of the Rashba parameter: $\alpha^{(1)}_\rm{R} = 0.02~\mathrm{a.u.}$ (blue) and $\alpha^{(2)}_\rm{R} = 0.04~\mathrm{a.u.}$ (magenta), used to analyze the peak position dependence on the spin-orbit coupling strength. Vertical dashed lines in panel (a) indicate the value of $k_{\mathrm{F}+} + k_{\mathrm{F}-}$ for each $\varepsilon_\rm{F}$, while in panel (b), they mark the location of the peaks sensitive to~$\alpha_\rm{R}$.}
\end{figure}


\setcounter{equation}{0}
\setcounter{figure}{0}
\setcounter{table}{0}

\renewcommand{\theequation}{S\arabic{equation}}
\renewcommand{\thefigure}{S\arabic{figure}}
\renewcommand{\thetable}{S\arabic{table}}

\clearpage
\pagebreak
\widetext
\begin{center}
\textbf{\large Supplemental Material}
\end{center}

\section{I. Non-equilibrium treatment through the Keldysh formalism}

In this work, we study non-equilibrium steady states (NESS), which occur when the energy injected by the external drive is balanced out by the energy dissipating into the environment \cite{ikeda2020}. Therefore, our system needs to be coupled to the environment. For that purpose, we adopt the Keldysh formalism, and thus we write the non-equilibrium Green's function as the following $2\times 2 $ matrix:
\begin{equation}
\underline{G} = 
\begin{pmatrix}
G^\rm{R} & G^\rm{K} \\[4pt]
0   & G^\rm{A}
\end{pmatrix},
\end{equation}
where $G^\rm{R}$, $G^\rm{A}$ and $G^\rm{K}$ are the retarded, advanced and Keldysh Green's functions. Within this framework, the coupling between the system and the environment can be represented with a self-energy correction corresponding to a free-fermion bath modelled as
\begin{equation}
\underline{\mathstrut \Sigma_\rm{bath}} (\omega) = 
\begin{pmatrix}
- i \Gamma & -2i\Gamma F(\omega) \\[4pt]
0 & i \Gamma
\end{pmatrix}.
\end{equation}
Here $\Gamma$ indicates the spectral function of the bath, which can be assumed to be independent of $\omega$ as a simple treatment of the dissipation \cite{aoki2014}. The Keldysh component results from the fluctuation-dissipation relation, and $F(\omega)$ is a linear combination of the Fermi-Dirac distribution function, which is given by
\begin{equation}
F(\omega) = 1- 2 f(\omega) = \tanh \left( \frac{\beta (\omega - \varepsilon_\rm{F})}{2}\right)
\end{equation}
for fermions, and where $\beta$ is the inverse temperature with which the bath is in equilibrium and $\varepsilon_\rm{F}$ the Fermi energy of the system. Now, so as to combine this with the Floquet formalism, the previous distribution function is rewritten~as
\begin{equation}
F_{mn}(\omega) =  \tanh \left( \frac{\beta (\omega + m\omega_0 - \varepsilon_\rm{F})}{2}\right) \delta_{mn}.
\end{equation}

With this, the non-equilibrium Green's function can be calculated through the following Dyson equation:
\begin{equation} \label{dyson_neq}
\underline{G}^{-1} = \underline{\mathstrut G_\rm{eq}}^{-1}  - \underline{\mathstrut \Sigma_\rm{bath}} ,
\end{equation}
which we write in this particular form for convenience. In our calculation process, the retarded component of the equilibrium Green's function is the direct solution of the Dyson equation (Eq. \ref{dyson} in main text), and the advanced one is its conjugate transpose, 
\begin{equation} \label{G^A_no-spin}
G_{\rm{eq},mn}^\mathrm{A}(\bm{r}, \bm{r}', \omega) = G_{\rm{eq},nm}^{\rm{R}*}(\bm{r}', \bm{r}, \omega).
\end{equation}
Regarding the Keldysh component, it can be assumed that its inverse vanishes as the equilibrium system is dissipationless \cite{aoki2014}, and hence Equation \ref{dyson_neq} becomes
\begin{equation} \label{dyson_neq2}
\begin{pmatrix}
G^\rm{R} & G^\rm{K} \\[4pt]
0        & G^\rm{A}
\end{pmatrix} = 
\begin{pmatrix}
\left( G_\rm{eq}^\rm{R}  \right)^{-1}  + i \Gamma & 2i\Gamma F  \\[4pt]
0 & \left( G_\rm{eq}^\rm{A}\right)^{-1} -i\Gamma &
\end{pmatrix}^{-1},
\end{equation}
where we omit the $\omega$ dependency for brevity. Note that, this way, the non-equilibrium Green's function can be calculated through the generalized formula of inversion of upper triangular block matrices \cite{meyer1970}, which does not require us to invert the whole right-hand side matrix, only one of its diagonals (retarded or advanced component). Likewise, it is possible to compute this inverse through the so-called Woodbury matrix identity \cite{henderson1981}, in such a way that direct inversion is avoided. Once the retarded, advanced and Keldysh components are calculated, we obtain the lesser Green's function through the $G^< \equiv (G^\rm{K} - G^\rm{R} + G^\rm{A})/2$ definition.

Finally, note that the equations above use scalar Green's functions. If we take into account the fact that our Green's functions are $2 \times 2$ spinors, the advanced component is calculated as
\begin{equation}
\bm{G}_{\mathbf{eq},\bm{mn}}^\mathbf{A}(\bm{r}, \bm{r}', \omega) = \left[ \bm{G}_{\mathbf{eq},\bm{nm}}^\mathbf{R}(\bm{r}', \bm{r}, \omega) \right]^\dagger,
\end{equation}
and the rest of the calculation process is carried out accordingly.


\section{II. Rashba Hamiltonian and calculation parameters}

\begin{figure}[h]
\includegraphics[width=11.8cm]{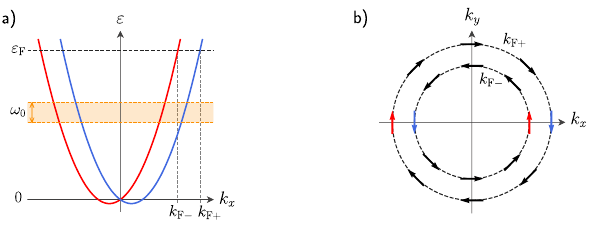}
\caption{Representation of Rashba Hamiltonian and system parameters: $m^*=1 \; \rm{a.u.}$, $\alpha_\rm{R} = 0.04 \; \rm{a.u.}$, $\varepsilon_\rm{F} = 0.03\;\rm{a.u.}$ and $\omega_0 = 0.004 \;\rm{a.u.}$. \textbf{a)} Dispersion relation of spin up (red) and down (blue) bands. $\varepsilon_\rm{F}$ is the Fermi energy and the orange shaded area illustrates an arbitrary frequency Brillouin zone, whose height is given by the fundamental frequency $\omega_0$. \textbf{b)} Rashba bands at the Fermi level, given by two concentric circles of radii $k_{\rm{F}+}$ and $k_{\rm{F}-}$. The arrows depict the spin orientation, showcasing the spin-momentum locking of the Hamiltonian. \label{params}}
\vspace{0.2cm}
\end{figure}

The Rashba Hamiltonian is 
\begin{equation} \label{rashba}
H = H_0 + H_\rm{R} = \frac{\bm{p}^{2}_\parallel}{2m^*} + \alpha_\rm{R}(\bm{\sigma} \times \bm{p}_\parallel),
\end{equation}
where we use atomic units, $m^*$ is the effective mass, $\alpha_\rm{R}$ is the Rashba parameter and $\bm{\sigma} = (\sigma_x, \sigma_y)$ are Pauli matrices indicating spin orientation. The resulting eigenvalues are given by
\begin{equation} \label{e_rashba}
\varepsilon_{\bm{k} \sigma} = \frac{\bm{k}^2}{2 m^*} + \sigma \alpha_\rm{R} \norm{\bm{k}},
\end{equation}
where $\sigma = \pm 1$ indicates the spin state, and a schematic representation of the corresponding spin up ($\sigma = +1$) and down ($\sigma = -1$) bands is displayed in Figure \ref{params}.

We consider the following parameters for our main calculation: $m^*=1 \; \rm{a.u.}$, $\alpha_\rm{R} = 0.04 \; \rm{a.u.}$, $\varepsilon_\rm{F} = 0.03\;\rm{a.u.}$ and $\omega_0 = 0.004 \;\rm{a.u.}$ (corresponding to a linear frequency of $\nu_0\sim 25\;\rm{THz}$); the representations in Figure \ref{params} are calculated according to these values. Here we should note that, even though the chosen parameters do not represent any system in particular, they have been selected to approximate typical ranges found for the magnitudes under consideration~\cite{bihlmayer2015}. Regarding the cylindrical potential, previous works showed that this kind of model potentials achieve good agreement with experiments for $v_0 = 0.1\; \rm{a.u.}$ \cite{harbury1996, rahachou2004}, and our tests showed that a small radius of $r_V=2.5\;\rm{a.u.}$ was sufficient for a proper visualization of the effect; besides, $\varphi = 0$ is considered for the cosine function of the time-dependence. Regarding the non-equilibrium aspect, we opt for a weak system-bath coupling of $\Gamma = 10^{-5} \;\rm{a.u.}$ and a low temperature limit of $\beta \rightarrow \infty$.


\section{III. Calculation of the unperturbed Green's function}

We obtain the two-dimensional, one-particle, unperturbed Green's function through the following definition:
\begin{equation}
\mathbf{G^0} (\bm{r}, \bm{r}', \omega) = \frac{1}{(2\pi)^2} \sum_\sigma \int \dd \bm{k} \; \frac{\psi_{\bm{k}\sigma}(\bm{r})\, \psi^{*}_{\bm{k}\sigma}(\bm{r}')}{\omega - \varepsilon_{\bm{k}\sigma} + i \eta},
\end{equation}
where $\psi_{\bm{k}\sigma}$ is the one-particle wave function, $\varepsilon_{\bm{k} \sigma}$ the corresponding energy, $\sigma $ represents the spin state and $\eta$ is an infinitesimal number; the wave functions are
\begin{equation}
\psi_{\bm{k}\sigma} (\bm{r}) = \frac{\mathrm{e}^{i\bm{k} \cdot \bm{r}}}{\sqrt{2}} \begin{pmatrix}
1 \\[4pt]
-i \sigma \mathrm{e}^{i\theta} 
\end{pmatrix},
\end{equation}
where $\theta$ describes the orientation of the spin in the 2D plane. With this, we have
\begin{equation}
\bm{G^0} (\bm{r}, \bm{r}', \omega) = \frac{1}{(2\pi)^2} \sum_\sigma \int \, \dd \bm{k} \: \frac{\mathrm{e}^{i\bm{k}\cdot(\bm{r}-\bm{r}')}}{2} \: \frac{
\begin{pmatrix}
1 & i\sigma \mathrm{e}^{-i\theta} \\[4pt]
-i\sigma\mathrm{e}^{i\theta} & 1
\end{pmatrix}
}{\omega - \varepsilon_{\bm{k}\sigma} +  i\eta}.
\end{equation}
Now we decompose the $\dd \bm{k}$ differential as $\dd \bm{k} = k\,\dd k\, \dd \theta$, and we get
\begin{align} \label{g0_int_decomp}
\bm{G^0} (\bm{r}, \bm{r}', \omega) = \frac{1}{8\pi^2} \sum_\sigma \int_{0}^{\infty} \dd k \, \frac{k}{ \omega -\varepsilon_{k \sigma} + i\eta} \: \int_{0}^{2\pi} \dd \theta \: \mathrm{e}^{i k \lVert \bm{r}-\bm{r}' \rVert \cos (\theta - \theta_{\bm{r}\bm{r}'})} \begin{pmatrix}
1 & i\sigma \mathrm{e}^{-i\theta} \\[4pt]
-i\sigma\mathrm{e}^{i\theta} & 1
\end{pmatrix},
\end{align}
where $\theta_{\bm{r}\bm{r}'}$ represents the angle of the $\bm{r}-\bm{r}'$ vector, that is,
\begin{equation} \label{exp}
\mathrm{e}^{\pm i \theta_{\bm{r}\bm{r}'}} = \cos \theta_{\bm{r}\bm{r}'} \pm i \sin \theta_{\bm{r}\bm{r}'} = \frac{(r_x - r_x') \pm i(r_y - r_y' )}{\norm{\bm{r}-\bm{r}'}}.
\end{equation}
In Equation \ref{g0_int_decomp} we used the fact that the $\varepsilon_{\bm{k} \sigma}$ eigenvalues of the Rashba Hamiltonian depend on $\bm{k}$ only through its modulus, as we can observe in Equation \ref{e_rashba}.

The angular part of the integrals is related to Bessel functions as
\begin{align}
& \int_{0}^{2\pi} \dd \theta \: \mathrm{e}^{i k \lVert \bm{r}-\bm{r}' \rVert \cos (\theta - \theta_{\bm{r}\bm{r}'})} = 2 \pi  J_0 (k\norm{\bm{r}-\bm{r}'}), \label{bessel0}\\[4pt]
& \int_{0}^{2\pi} \dd \theta \: \mathrm{e}^{i k \lVert \bm{r}-\bm{r}' \rVert \cos (\theta - \theta_{\bm{r}\bm{r}'})} \,\mathrm{e}^{\pm i\theta} = 2\pi i  \mathrm{e}^{\pm i\theta_{\bm{r}\bm{r}'}}  J_1 (k\lVert \bm{r}-\bm{r}' \rVert); \label{bessel1}
\end{align}
and, therefore, we have
\begin{align}
G_{00}^{0} (\bm{r}, \bm{r}', \omega) &= \frac{1}{4\pi} \sum_\sigma \int_{0}^{\infty} \dd k \: \frac{\,k J_0 (k \norm{\mathbf{r}-\mathbf{r'}})\,}{\omega -\varepsilon_{k\sigma} +  i\eta}, \\[4pt]
G_{01}^{0} (\bm{r}, \bm{r}', \omega) &= -\frac{\mathrm{e}^{-i \theta_r}}{4\pi} \sum_\sigma \int_{0}^{\infty} \dd k \: \frac{\, \sigma k J_1 (k\norm{\mathbf{r}-\mathbf{r}'})\,}{ \omega -\varepsilon_{k\sigma} + i\eta}, \\[4pt]
G_{10}^{0} (\bm{r}, \bm{r}', \omega) &= \frac{\mathrm{e}^{i \theta_r}}{4\pi} \sum_\sigma \int_{0}^{\infty} \dd k \: \frac{\, \sigma k  J_1 (k\norm{\mathbf{r}-\mathbf{r}'})}{ \omega -\varepsilon_{k\sigma} + i\eta},
\end{align}
for the diagonal (note that $G_{11}^0$ = $G_{00}^0$) and off-diagonal components of the unperturbed Green's function. Regarding the radial part, we first define $G_\rm{d}^0 \equiv G_{00}^0$ and
\begin{equation}
G^{0}_{\mathrm{od}} (\bm{r}, \bm{r}', \omega) \equiv \frac{1}{4\pi} \sum_\sigma \int_{0}^{\infty} \dd k \: \frac{\, \sigma k J_1 (k\norm{\bm{r}-\bm{r}'})}{\omega -\varepsilon_{k\sigma} +  i\eta}.
\end{equation}
In order to compute these, we rewrite the denominator by using the so-called Sokhotski-Plemelj formula \cite{economou}, 
\begin{equation} \label{sok-plem}
\lim_{\eta \rightarrow 0} \, \frac{1}{ \omega -\varepsilon_{k\sigma}  \pm i\eta} = \mathcal{P.V.} \Big( \frac{1}{\omega -\varepsilon_{k\sigma}} \Big) \mp i \pi \delta \left(\omega -\varepsilon_{k\sigma}\right),
\end{equation}
where $\mathcal{P.V.}$ denotes the Cauchy principal value and $\delta$ the Dirac delta, and then we take its imaginary part, that is,
\begin{align}
\Im G_{\mathrm{d}}^{0} (\bm{r}, \bm{r}', \omega)  &= -\frac{1}{4} \sum_\sigma \int_{0}^{\infty} \dd k\: k J_0(k\norm{\bm{r}-\bm{r}'}) \, \delta(\omega -\varepsilon_{k\sigma}), \label{im_g0d}\\[4pt]
\Im G_{\mathrm{od}}^{0} (\bm{r}, \bm{r}', \omega) &= -\frac{1}{4}  \sum_\sigma \int_{0}^{\infty} \dd k\: \sigma k J_1(k \norm{\bm{r}-\bm{r}'})\, \delta(\omega -\varepsilon_{k\sigma}). \label{im_g0od}
\end{align}
Due to the Dirac delta, the result of the integrals above is a Bessel function ($J_0$ for the diagonal component, $J_1$ for the off-diagonal) multiplied by a set of parameters. With this, we can obtain the real part through the Hilbert transform, which describes the relationship between the real and the imaginary part of an analytic function. In the case of the Bessel function of the first kind, one obtains
$\mathcal{H} J_{n} (x) \approx -Y_{n} (x)$ \cite{poularikas}, where $n=0,1$ and $Y$ is the Bessel function of the second kind. At this point, it is useful to recall that Hankel functions of the first kind are given by $H_{n}^{(1)} = J_n + i Y_n$. In this manner, we can write the radial integrals as
\begin{align}
& G_\rm{d}^0 (\bm{r}, \bm{r}',\omega) = -\frac{im^*}{4} \sum_\sigma A_\sigma H_0^{(1)} \left( a_\sigma m^*\norm{\bm{r}-\bm{r}'} \right), \\
& G_\rm{od}^0 (\bm{r}, \bm{r}',\omega) = -\frac{im^*}{4} \sum_\sigma \sigma A_\sigma H_1^{(1)} \left( a_\sigma m^* \norm{\bm{r}-\bm{r}'} \right),
\end{align}
where the parameters $A_\sigma$ and $a_\sigma$ depend on $\omega$ and the Rashba parameter $\alpha_\rm{R}$, as $A_\sigma \equiv 1 + \sigma \alpha_\rm{R}/ \sqrt{\alpha_\rm{R}^2 + 2 \omega}$ and $a_\sigma \equiv \sqrt{\alpha_\rm{R}^ 2 + 2 \omega} + \sigma \alpha_\rm{R}$. Hence, the final unperturbed Green's function is given by
\begin{equation}
\bm{G^0} (\bm{r}, \bm{r}',\omega) = 
\begin{pmatrix}
G_\rm{d}^0 (\bm{r}, \bm{r}',\omega) & -\ee^{-i\theta_{\bm{r}\bm{r}'}} G_\rm{od}^0 (\bm{r}, \bm{r}',\omega) \\[4pt]
\ee^{i\theta_{\bm{r}\bm{r}'}} G_\rm{od}^0 (\bm{r}, \bm{r}',\omega) & G_\rm{d}^0 (\bm{r}, \bm{r}',\omega)
\end{pmatrix}.
\end{equation}


\section{IV. Calculation of mean square deviation}

We define the time average of the mean square deviation of the (normalized) magnetization over one period as
\begin{equation}
\langle\!\langle \tilde{m}_{x,y} \rangle \!\rangle_T \,(\bm{r}) = \sqrt{\langle \tilde{m}_{x,y}^2 (\bm{r}, t) \rangle_T - \langle \tilde{m}_{x,y}(\bm{r}, t) \rangle_T^2},
\end{equation}
where angle brackets indicate time averages over one period $T$ and where, in our case, $\langle \tilde{m}_{x,y}(\bm{r}, t) \rangle_T$ is zero for all $\bm{r}$ coordinates.


\section{V. Investigation of parameter dependence}

The calculations of the main text of this letter are performed for a single set of parameters, as presented in Section II of this Supplemental Material. Nevertheless, it can be interesting to study the dependence of the induced magnetization on the key external drive parameters, namely the fundamental frequency $\omega_0$ and the oscillation amplitude $v_0$, as it can provide more information on the tunability of the phenomenon. 

\begin{figure}[h!]
\begin{minipage}[h!]{0.45\linewidth}
\centering
	\includegraphics[width=0.75\linewidth]{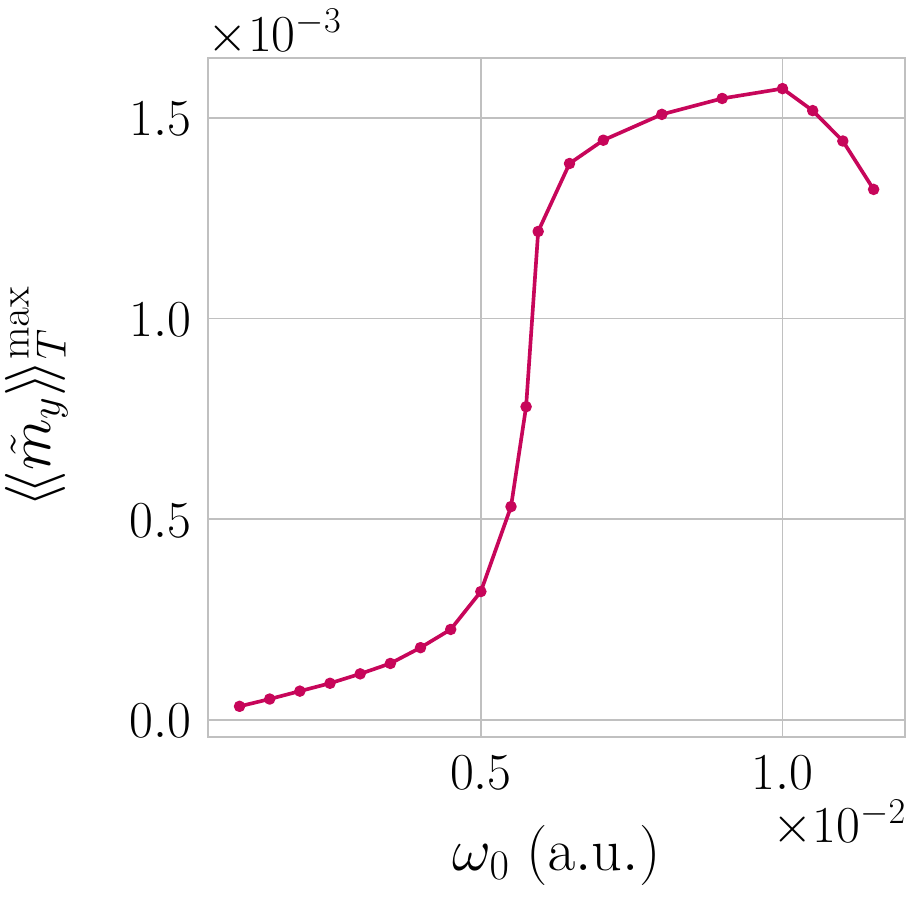}
	\vspace{-0.2cm}
	\caption{Maximum value of $\langle\!\langle \tilde{m}_{y} \rangle \!\rangle_T \,(\bm{r})$ as a function of the fundamental frequency $\omega_0$. We use $\omega_0 = 0.004$ a.u. in the calculations of the main text.}
	\label{w0dep}
\end{minipage}
\hspace{1.cm}
\begin{minipage}[h!]{0.45\textwidth}
\centering
	\includegraphics[width=0.75\linewidth]{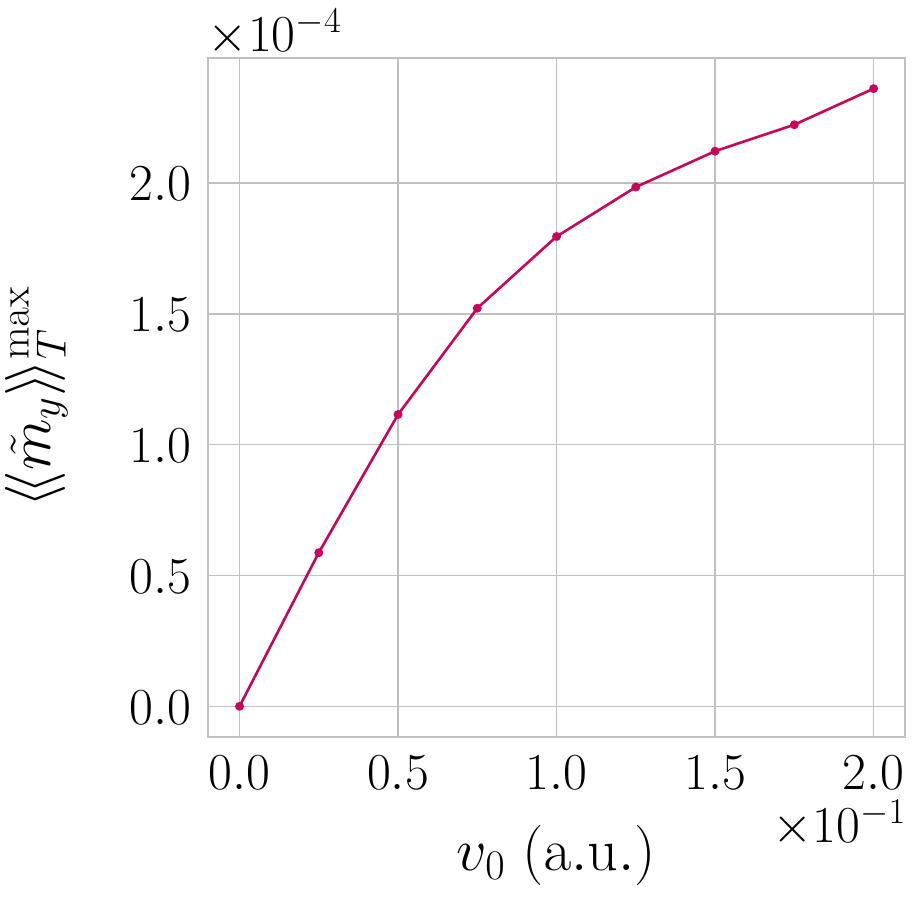}
	\vspace{-0.2cm}
	\caption{Maximum value of $\langle\!\langle \tilde{m}_{y} \rangle \!\rangle_T \,(\bm{r})$ as a function of the drive oscillation amplitude $v_0$. We use $v_0 = 0.1$ a.u. in the calculations of the main text.}
	\label{v0dep}
\end{minipage}
\end{figure}

Hence, in Figures \ref{w0dep} and \ref{v0dep} we present the maximum value of the time-averaged mean square deviation $\langle\!\langle \tilde{m}_{y} \rangle \!\rangle_T \,(\bm{r})$ as a function of $\omega_0$ and $v_0$, respectively. We opt for this quantity as it provides insight into the strength of the phenomenon through a single time-averaged number. 
We observe a linear increase of the magnetic response with $\omega_0$ and $v_0$, a slow growth for increasing values of $v_0$, while it reaches a maximum in $\omega_0$  at frequencies of the order of the characteristic spin-orbit splitting energy scale, which is $\alpha_\mathrm{R} k_\mathrm{F}$ for a Rashba system.


\section{VI. Momentum resolved Green's functions at different times}

\begin{figure}[h!]
\begin{minipage}[h!]{0.47\textwidth}
  \centering
  \includegraphics[width=1\linewidth]{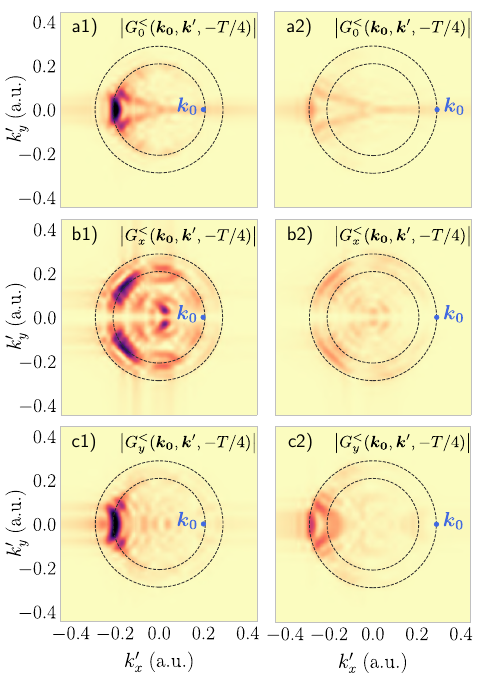}
  \caption{Absolute value of momentum resolved lesser Green's functions $G^{<}_0$ (a), $G^{<}_x$ (b) and $G^{<}_y$ (c) evaluated at time $t=-T/4$. Initial momentum $\bm{k_0}$ is located close to the inner (left-hand side panels (a1), (b1) and (c1)) and outer (right-hand side panels (a2), (b2) and (c2)) Fermi contours.}
  \label{2dfft_t4}
\end{minipage}
\hspace{0.6cm}
\begin{minipage}[h!]{0.47\textwidth}
  \centering
  \includegraphics[width=1\linewidth]{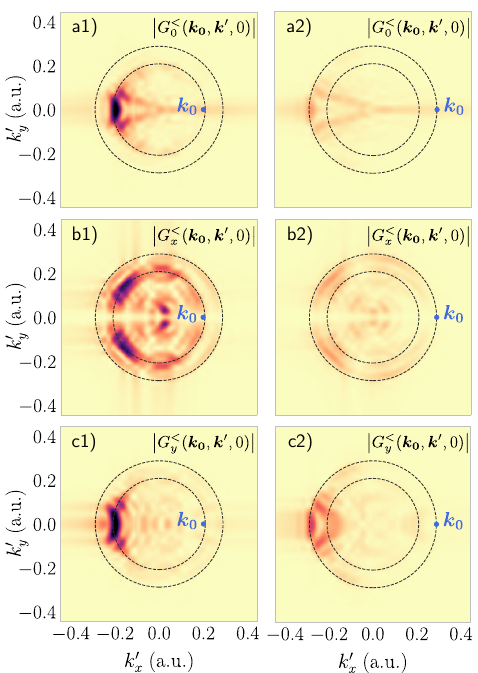}
  \caption{Absolute value of momentum resolved lesser Green's functions $G^{<}_0$ (a), $G^{<}_x$ (b) and $G^{<}_y$ (c) evaluated at time $t=0$. Initial momentum $\bm{k_0}$ is located close to the inner (left-hand side panels (a1), (b1) and (c1)) and outer (right-hand side panels (a2), (b2) and (c2)) Fermi contours.}
  \label{2dfft_t0}
\end{minipage}
\end{figure}

The main text of this letter presents the snapshot at $t = -T/2$ as representative. In Figures~\ref{2dfft_t4} and~\ref{2dfft_t0}, we show the momentum resolved lesser Green's functions at $t = -T/4$ and $t = 0$, respectively, to demonstrate that in momentum space this picture remains nearly unchanged, the phenomenon is robust, and in fact, any time could be considered representative.

\end{document}